\documentclass[showpacs,preprintnumbers,amsmath,amssymb,prb,floatfix]{revtex4}
\usepackage{graphicx}
\usepackage{dcolumn}
\usepackage{bm}
\usepackage{amsmath,amsthm,amssymb,ascmac}


\newcommand{\be}{\beta}
\newcommand{\ka}{\kappa}
\newcommand{\bea}{\begin{eqnarray}}
\newcommand{\eea}{\end{eqnarray}}
\newcommand{\aeq}{\!\! &=& \!\!}

\newcommand{\mL}{\mathcal{L}}

\newcommand{\mH}{\mathcal{H}}
\newcommand{\mD}{\mathcal{D}}
\newcommand{\D}{{\rm{D}}}
\newcommand{\G}{{\rm{G}}}

\newcommand{\dg}{^\dagger}
\newcommand{\tl}{\tilde}

\newcommand{\defe}{\stackrel{\mathrm{def}}{=}}
\newcommand{\e}{\equiv}
\newcommand{\ep}{\varepsilon}
\newcommand{\ga}{\gamma}

\newcommand{\al}{\alpha}

\newcommand{\f}{\frac}
\newcommand{\half}{\frac{1}{2}}
\newcommand{\pr}{\prime}
\newcommand{\dl}{\delta}
\newcommand{\Dl}{\Delta}
\newcommand{\lm}{\lambda}
\newcommand{\Lm}{\Lambda}

\newcommand{\om}{\omega}
\newcommand{\Om}{\Omega}

\newcommand{\tot}{{\rm{tot}}}

\newcommand{\com}{\hspace{0.5mm}, \quad}
\newcommand{\la}{\label}

\newcommand{\no}{\nonumber}

\newcommand{\re}[1]{(\ref{#1})}

\newcommand{\res}[1]{˜\ref{#1}}

\newcommand{\RM}[1]{{\rm{#1}}}
\newcommand{\p}{{\varphi}}
\newcommand{\w}{\wedge}
\newcommand{\co}[1]{``{#1}''}
\newcommand{\Ref}[1]{Ref.[\onlinecite{#1}]}

\begin{document}

\title{Application of covariant analytic mechanics with differential forms to gravity with Dirac field}

\author{Satoshi Nakajima}
\email{subarusatosi@gmail.com}

\date{\today}

\affiliation{
Graduate School of Pure and Applied Sciences, University of Tsukuba, 
1-1-1, Tennodai, Tsukuba, Japan 305-8571}
\date{\today}

\begin{abstract}
We apply the covariant analytic mechanics with the differential forms to the Dirac field and the gravity with the Dirac field.
The covariant analytic mechanics treats space and time on an equal footing regarding the differential forms as the basic variables.
A significant feature of the covariant analytic mechanics is that the canonical equations, in addition to the Euler-Lagrange equation, are not only
manifestly general coordinate covariant but also gauge covariant. 
Combining our study and the previous works (the scalar field, the abelian and non-abelian gauge fields 
and the gravity without the Dirac field), the applicability of the covariant analytic mechanics is checked for all fundamental fields. 
We study both the first and second order formalism of the gravitational field coupled with matters including the Dirac field. 
It is suggested that gravitation theories including higher order curvatures cannot be treated by the second order formalism in the covariant analytic mechanics.
\end{abstract}

\maketitle

\section{Introduction}
In the traditional analytic mechanics, the Hamilton formalism gives especial weight to time.
Then, the Lorentz covariance is not trivial. 
Moreover, for the constrained system, for instance the gauge field, the gauge fixing or the Dirac's theory is needed.
The De Donder-Weyl theory solves the former problem \cite{Weyl,IV,IV2}. 
In this theory, the conjugate generalized momenta of a field $\psi$ are introduced as $\pi^\mu=\partial \mL/\partial (\partial_\mu \psi)$ $(\mu=0,1,2,3)$, 
where $\mL$ is the Lagrangian density.
$\pi^0$ is the traditional conjugate momentum.
The generalization of the Hamiltonian density is given by $\mH(\psi,\pi^\mu)=\partial_\mu \psi \pi^\mu-\mL$.
The common property of the traditional analytic mechanics and the De Donder-Weyl theory is that its basic variables of 
the variation are components of the tensor.

By the way, the Lagrange formalism is sometimes formulated by the differential forms \cite{75K,75T,80,82,95}, 
in which the basic variables are the differential forms.
Because the differential form is independent of the coordinate system, the general coordinate covariance is guaranteed manifestly.
And this formulation often largely reduces the cost of the calculation. 
Nakamura \cite{Na} generalized this formulation to the Hamilton formalism. 
In this method, the conjugate momentum is also a differential form, which treats space and time on an equal footing. 
Nakamura applied this method to the Proca field and the electromagnetic field with manifest covariance and, in the latter, with the gauge covariance. 
The conjugate momentum form becomes independent degree of freedoms. 
Kaminaga \cite{K} formulated strictly mathematically Nakamura's idea and constructed the general theory in arbitrary dimension.
We called this theory as the {\it covariant analytic mechanics}.
Kaminaga studied that the Newtonian mechanics of a harmonic oscillator ($(0+1)$ dimension) and the scalar field, 
 the abelian and non-abelian gauge fields and 4 dimension gravity without the Dirac field.
The gravitational field was formulated by the  second order formalism, in which the basic variable is only the frame (vielbein). 
And the absence of the torsion was assumed. 
On the other hand, Nester \cite{N04} also investigated independently the covariant analytic mechanics and 
 constructed the general theory in 4 dimension and applied it to the Proca field, the electromagnetic field and the non-abelian gauge field.
As we will explain in \res{1st}, the treatment of the gravitational field of Nester was not complete Hamilton formalism although 
the  conjugate momentum forms of the frame and the connection were introduced. 
The original idea was mentioned in \Ref{N91}.

We investigate the Kaminaga's study. 
We apply the covariant analytic mechanics to the Dirac field and the gravity with the Dirac field for the first time.
Combining our study and the previous works, the applicability of the covariant analytic mechanics is checked for all fundamental fields. 
It is suggested that gravitation theories including higher order curvatures can not be treated by the second order formalism in the covariant analytic mechanics.

In \res{CM}, we review the covariant analytic mechanics with the application to the  electromagnetic field and 
introduce the Poisson bracket for first time.
In \res{Notation}, we explain the several notations and in \res{Dirac field} we study the Dirac field.
In \res{GD}, we study the gravitational field coupled with matters including the Dirac field.
First, we discuss the first order formalism, in which both the frame and the connection are basic variables (\res{1st}).
Next, we move to the second order formalism.
In \res{L_F}, the Lagrange formalism is studied and we show that the Lagrange form of the pure gravity is given by subtracting the total differential term 
from the Einstein-Hilbert form.
We discuss that if we do not drop the total differential term, it is probably impossible to derive the correct equations.
In \res{H_F}, we move to the Hamilton formalism and give a broad overview of the remainder discussion.
In \res{H_V}, we take the derivatives of the Hamilton form of the pure gravity using specialties of 4 dimension system and in \res{C_eq}, 
we discuss the canonical equations. 

\section{Covariant analytic mechanics} \la{CM}

\subsection{General theory} \la{General}

Let us consider $D$ dimension space (pseudo-Riemannian or Riemannian space). 
Suppose a differential $p$-form $\be$ ($p=0,1,\cdots,D$) is described by differential forms $\{\al^i \}_{i=1,\cdots,k}$. 
If there exists the differential form $\om_i$ such that $\be$ behaves under variations $\dl \al^i$ as 
\bea
\dl \be = \sum_i \dl \al^i \wedge \om_i ,
\eea
we call $\om_i$ the {\it derivative} of $\be$ by $\al^i$ and denote 
\bea
\f{\partial \be}{\partial \al^i} \defe  \om_i,
\eea
namely, $\dl \be \e \sum_i \dl \al^i \wedge \f{\partial \be}{\partial \al^i}$. 
If $\al^i$ is $q_i(\le p)$-form, $\partial \be/\partial \al^i$ is $(p-q_i)$-form.

As the traditional analytic mechanics starts from the Lagrangian density, the covariant analytic mechanics starts from 
{\it Lagrange $D$-form} $L$. 
$L$ is a function of $\psi$ and $d\psi$, $L=L(\psi,d\psi)$, where $\psi$ is a set the differential forms.
For simplicity, we treat $\psi$ as single $p$-form. 
The variation of $L$ is given by
\bea
\dl L = \dl \psi \wedge \f{\partial L}{\partial \psi}+\dl d \psi \wedge \f{\partial L}{\partial d\psi}.
\eea
Since the second term of RHS can be rewritten as
\bea
\dl d \psi \wedge \f{\partial L}{\partial d\psi} = d\Big(\dl \psi \wedge \f{\partial L}{\partial d\psi} \Big)-(-1)^p\dl \psi \wedge d\f{\partial L}{\partial d\psi} ,
\eea
we obtain
\bea
\dl L = \dl \psi \wedge \Big( \f{\partial L}{\partial \psi}-(-1)^p d\f{\partial L}{\partial d\psi} \Big)
+d\Big(\dl \psi \wedge \f{\partial L}{\partial d\psi} \Big).
\eea
Hence, the {\it Euler-Lagrange equation} is 
\bea
\f{\partial L}{\partial \psi}-(-1)^p d\f{\partial L}{\partial d\psi} = 0. \la{EL}
\eea

We define the {\it conjugate momentum form} $\pi$ as 
\bea
\pi \defe \f{\partial L}{\partial d\psi}.
\eea
$\pi$ is $D-p-1=q$-form. The {\it Hamilton $D$-form} (not $(D-1)$-form) is defined by
\bea
H =H(\psi,\pi)\defe d\psi \wedge \pi-L ,
\eea
as a function of $\psi$ and $\pi$. The variation of $H$ is given by
\bea
\dl H = (-1)^{(p+1)q}\dl \pi \wedge d\psi -\dl \psi \wedge\f{\partial L}{\partial \psi}.
\eea
Then, we obtain
\bea
\f{\partial H}{\partial \psi}=  -\f{\partial L}{\partial \psi}\com
\f{\partial H}{\partial \pi}= \ep_{p,D}d\psi, 
\eea
where $\ep_{p,D}\defe (-1)^{(p+1)q}=1$ if $p$ is an odd number and $\ep_{p,D}=-(-1)^D$ if $p$ is an even number.
By substituting the Euler-Lagrange equation \re{EL}, we obtain the {\it canonical equations} \cite{K}
\bea
d\psi  = \ep_{p,D}\f{\partial H}{\partial \pi} \com d\pi = -(-1)^p\f{\partial H}{\partial \psi}.
\eea
Now we introduce the {\it Poisson bracket} by
\bea
\{A,B\} \defe \ep_{p,D}\f{\partial A}{\partial \psi}\w \f{\partial B}{\partial \pi}-(-1)^p\f{\partial A}{\partial \pi}\w \f{\partial B}{\partial \psi}.
\eea
Then, the canonical equations can be written as
\bea
d\psi  =\{ \psi,H\} \com d\pi  =\{ \pi,H\}.
\eea
And we have 
\bea
\{\psi,\pi\}=\ep_{p,D}\com \{\pi,\psi\}=-(-1)^p.
\eea
The applicability of the Poisson bracket to the quantization is unclear. 
And the generalization of the canonical transform theory have not been studied.

Let consider $D$ dimension space-time, which has the metric $g_{\mu\nu}$. 
The Hodge operator $\ast$ maps an arbitrary $p$-form $\om=\om_{\mu_1 \cdots \mu_p}dx^{\mu_1}\w \cdots \w dx^{\mu_p}$ $(p=0,1,\cdots,D)$ to $D-p=r$-form as
\bea
\ast \om = \f{1}{r!}E_{\nu_1 \cdots \nu_r}^{\ \ \ \ \ \ \mu_1 \cdots \mu_p}\om_{\mu_1 \cdots \mu_p}dx^{\nu_1} \w \cdots \w dx^{\nu_r}.
\eea
Here, $E_{\mu_1 \cdots \mu_D}$ is the complete anti-symmetric tensor such that $E_{01\cdots D-1}=\sqrt{-g}$ ($g=\RM{det} g_{\mu\nu}$). 
And $\ast \ast \om=-(-1)^{p(D-p)}\om$ holds. In particular, $\Om=\ast 1$ is the volume form. 
The Lagrange form $L$ relates to the Lagrangian density $\mL$ as $L=\mL \Om$. 
In the following of this section, we set $D=4$. Then, $\ast \ast \om=-(-1)^p\om$ holds. 
The Lagrangian density $\mL_\mD$ corresponding to a Lagrange form $d\mD$ is given by $\mL_\mD \sqrt{-g}=\partial_\mu (\sqrt{-g}d^\mu)$, 
where $\mD$ is a 3-form $\mD=d_{\al\be\ga}dx^\al \w dx^\be \w dx^\ga $ and $d^\mu = d_{\al\be\ga} E^{\al\be\ga\mu}$.
If $\mD$ dose not include $d\psi$, $d\mD$ dose not contribute to the Euler-Lagrange equation.
We will discuss an instance of $\mD$ including $d\psi$ in \res{L_F}. 
The Lagrange form corresponding to a Lagrangian density $a_\mu b^\mu$ is given by $\ast a \w b $ with $a=a_\mu dx^\mu$ and $b=b_\mu dx^\mu$.
And the Lagrange form corresponding to a Lagrangian density $\half c_{\mu\nu}d^{\mu\nu}$ is given by $ \ast c \w d=c \w \ast d$ 
with $c=\half c_{\mu\nu}dx^\mu \w dx^\nu$ and $d=\half d_{\mu\nu}dx^\mu \w dx^\nu$.

\subsection{Electromagnetic field} \la{EM}

The Lagrange form of the electromagnetic field is given by
\bea
L =L(A,dA)= -\f{1}{2}F\w \ast F+J \w A=\mL \Om ,
\eea
where
\bea
\mL =\mL(A_\mu,\partial_\mu A_\nu) = -\f{1}{4}F_{\mu\nu}F^{\mu\nu}+A_\mu J^\mu ,
\eea
$A=A_\mu dx^\mu$, $J=\ast (J_\mu dx^\mu)$ and $F=dA=\half F_{\mu\nu}dx^\mu \w dx^\nu$ with $F_{\mu\nu}=\partial_\mu A_\nu-\partial_\nu A_\mu$.
$A_\mu$ is the vector potential and  $J^\mu$ is the current density, which is independent of $A_\mu$. 
We obtain $\partial L/\partial A=-J$ and $\partial L/\partial dA=-\ast F$ using $\dl \ast F=\ast \dl F $.
The Euler-Lagrange equation $\partial L/\partial A+d(\partial L/\partial dA)=0$ is 
\bea
d\ast F=-J \la{Maxwell} . 
\eea
This equation and an identity $dF=0$ are the Maxwell equations. 

In contrast to that the basis variables of the variation of the traditional analytic mechanics are components of the tensor, $A_\mu$, 
the basis variable (reality) is the differential form $A=A_\mu dx^\mu$ in the covariant analytic mechanics. 
In general, the Lagrange formalism is equivalent to the traditional one. 
However, as we show just after, the Hamilton formalism is not equivalent to the traditional one.

In the traditional analytic mechanics, the definition of the conjugate momentum, \\
$\Pi^\mu=\partial \mL/\partial (\partial_0A_\mu)=-F^{0\mu}$, gives especial weight to time. 
Then, the Lorentz covariance is not trivial.  
Moreover, because $\Pi^0=0$, this system is a constrained system, which needs to the gauge fixing or the Dirac's theory (Dirac bracket).
In contrast, the Hamilton formalism of the covariant analytic mechanics is manifestly Lorentz covariant since 
the differential forms are independent of the coordinate system.
Moreover, the conjugate momentum form, $\pi=\partial L/\partial dA=-\ast F$, can represent $dA$ as $dA=F=\ast \pi$.
So, the gauge fixing or the Dirac's theory are not needed. 
This formulation is gauge free. 
The position variable is a 1-form, which has 4 components, and the conjugate momentum variable is a 2-from, 
which has 6 components (electric and magnetic fields).

The Hamilton form is given by
\bea
H(A,\pi) = \half \pi \w \ast \pi-J\w A.
\eea
We have $\partial H/\partial \pi=\ast \pi$ and $\partial H/\partial A=J$. 
The canonical equations $dA=\partial H/\partial \pi$ and $d\pi=\partial H/\partial A$ are
\bea
dA=\ast \pi\com d\pi=J .
\eea
The former is equivalent to the definition of the conjugate momentum form and the latter 
coincides with the Euler-Lagrange equation \re{Maxwell}.

\section{Notation} \la{Notation}@
Let $g$ be the metric of which signature is $(-+\cdots+)$, and let $(\theta^a)$ denote an orthonormal frame.
$\theta^a$ can be expanded as $\theta^a=\theta^a_{\ \mu}dx^\mu$ with the vielbein $\theta^a_{\ \mu}$. 
We have $g=\eta_{ab}\theta^a \otimes \theta^b$ with $\eta_{ab}=(-+\cdots+)$.
We put $e^a =\ast \theta^a$, $e^{ab}=\ast(\theta^a \w \theta^b)$ and $e^{abc}=\ast(\theta^a \w \theta^b \w \theta^c)$.
Let $\om^a_{\ b}$ and $w^a_{\ b}$ respectively be the connection and the Levi-Civit\`{a} connection 1-form. 
All indices are lowered and raised with $\eta_{ab}$ or its inverse $\eta^{ab}$. 
Then, $\om_{ba}=-\om_{ab}$ and the first structure equation 
\bea
d\theta^a+\om^a_{\ b}\w \theta^b=\Theta^a , \la{FS}
\eea
hold. Here, $\Theta^a=\half C^a_{\ bc}\theta^b \w \theta^c$ is the torsion 2-form.
From \re{FS}, we obtain
\bea
\om_{abc}\aeq w_{abc}+\tl \om_{abc} ,\no\\
w_{abc} \aeq \half(\Dl_{cba}+\Dl_{abc}+\Dl_{bca}) \com \tl \om_{abc}=-\half(C_{cba}+C_{abc}+C_{bca}). \la{om_abc}
\eea
Here, we expanded $d\theta^a$ and $\om_{ab}$ as $d\theta^a=\half \Dl^a_{\ bc}\theta^b \w \theta^c$ and $\om_{ab}=\om_{abc}\theta^c$.
We have $w_{ab}=w_{abc}\theta^c$ and define $\tl \om_{ab}\defe \tl \om_{abc}\theta^c$,
$\om_a \defe \om^b_{\ ab}$ and $C_a\defe C^b_{\ ab}$.
In Appendix \ref{formula}, several identities about 
$\theta^a \w e_{a_1 \cdots a_r}$, $de_{a_1 \cdots a_r}$ $(r=1,2,3)$ and
$\dl e_{a_1 \cdots a_r}(r=0,1,2)$ are listed.
The curvature 2-form $\Om^a_{\ b}$ is given by $\Om^a_{\ b}=d\om^a_{\ b}+\om^a_{\ c}\w \om^c_{\ b}$.
Expanding the curvature form as $\Om^a_{\ b}=\half R^a_{\ bcd}\theta^c \w \theta^d$, we define 
$R_{ab}\defe R^c_{\ acb}$ and $R\defe R^a_{\ a}$.

In the following, we set $D=4$. 
However, up to \res{H_F}, the dimension dependence appears only in the sign factor of exchanging differential forms.
After \res{H_V}, we use specialties of $D=4$.

\section{Dirac field} \la{Dirac field}

\subsection{Lagrange formalism}

The Lagrange form of the Dirac field $\psi$ is given by
\bea
L_\D(\psi,\bar{\psi},d\psi,d\bar{\psi})
 = -\half \bar{\psi}\ga_c e^c \w (d\psi+\f{1}{4}\ga_{ab}\om^{ab}\psi)+\half e^c \w (d\bar{\psi}-\f{1}{4}\bar{\psi}\ga_{ab}\om^{ab})\ga_c\psi-m\bar{\psi}\psi \Om , \la{L_Ds}
\eea
with $\bar{\psi}= i\psi \dg \ga^0$ and $\ga_{ab}\defe \ga_{[a}\ga_{b]}$. 
Here, $\ga^a$ is the gamma matrix, which satisfies $\ga^{(a}\ga^{b)}=\eta^{ab}$.
( ) and [ ] are respectively the symmetrization and anti-symmetrization symbols. 
It is important that the frame (vielbein) is necessary to write down the Lagrange form even in the flat space-time. 
This is because that the Dirac field is a representation of the Lorentz transformation (not of the coordinate transformation) and 
a spinor field is defined in the tangent Minkowski space.
The connection $\om^{ab}$ is the gauge field for the local Lorentz transformations. 
Subtracting the total differential term $-\half d(e^a\bar{\psi}\ga_a\psi)$ from \re{L_Ds}, and using
\bea
 \ga_c  e^c \w \f{1}{4}\ga_{ab}\om^{ab}
\aeq  \f{1}{4} e^c \w \ga_{ab}\om^{ab}\ga_c+e^c \w \f{1}{4}[\ga_c,\ga_{ab}]\om^{ab} \no\\
\aeq \f{1}{4}  e^c \w \ga_{ab}\om^{ab}\ga_c+\ga^a \om_a \Om, \la{trick}
\eea
and \re{A10}, we obtain
\bea
L_\D^\pr(\psi,\bar{\psi},d\psi,d\bar{\psi}) = -\bar{\psi}\ga_c e^c \w (d+\f{1}{4}\ga_{ab}\om^{ab})\psi-m\bar{\psi} \psi\Om-\half C_a \bar{\psi}\ga^a\psi \Om \la{L_D} .
\eea
In the second line of \re{trick}, we used $\f{1}{4}[\ga_c,\ga_{ab}] =\f{1}{2}(- \eta_{bc}\ga_a+\eta_{ac}\ga_b)$ and $e_b \w \om_a^{\ b}=\om_a \Om$.
$C_a$ is regarded as independent of $\psi$ and $\bar{\psi}$. 
As we will show in \re{C_a=0}, $C_a=0$ is required. 
For simplicity, we treat the Dirac field as a usual number (not the Grassmann number).

From the variation by $\bar{\psi}$, we obtain 
\bea
\f{\partial L_\D^\pr}{\partial \bar{\psi}}=-\ga_c e^c \w (d+\f{1}{4}\ga_{ab}\om^{ab})\psi-m \psi\Om-\half C_a \ga^a\psi \Om \com \ \f{\partial L_\D^\pr}{\partial d\bar{\psi}}=0 .
\eea
The Euler-Lagrange equation $\partial L_\D^\pr/\partial \bar{\psi}-d(\partial L_\D^\pr/\partial d\bar{\psi})=0$ is given by
\bea
\ga_c e^c \w (d+\f{1}{4}\ga_{ab}\om^{ab})\psi+m \psi\Om+\half C_a \ga^a\psi \Om = 0.\la{DE}
\eea
This is equivalent to the Dirac equation. 
From the variation by $\psi$, we obtain 
\bea
\f{\partial L_\D^\pr}{\partial \psi}=-\bar{\psi}\ga_c e^c \w \f{1}{4}\ga_{ab}\om^{ab}-m\bar{\psi} \Om-\half C_a \bar{\psi}\ga^a \Om \com
 \ \f{\partial L_\D^\pr}{\partial d\psi}=\bar{\psi} \ga_a e^a .
\eea
The Euler-Lagrange equation $\partial L_\D^\pr/\partial \psi-d(\partial L_\D^\pr/\partial d\psi)=0$ is 
\bea
\bar{\psi}\ga_c e^c \w \f{1}{4}\ga_{ab}\om^{ab}+m\bar{\psi} \Om+\half C_a \bar{\psi}\ga^a \Om +d\bar{\psi}\w \ga_a e^a+\bar{\psi} \ga_a de^a = 0 .\la{Deq}
\eea
Using \re{trick} and \re{A10}, \re{Deq} becomes
\bea
(d\bar{\psi} -\f{1}{4} \bar{\psi}\ga_{ab} \om^{ab})\w \ga_c e^c+m\bar{\psi}\Om -\half C_a\bar{\psi}\ga^a  \Om = 0 .\la{DE2}
\eea
This is the Hermitian conjugate of \re{DE}.

\subsection{Hamilton formalism}

The conjugate momentum forms of $\psi$ and $\bar{\psi}$ are respectively  $\Pi=\bar{\psi} \ga_a e^a$ and $\bar{\Pi}=0$.
Then, the Hamilton form $H_\D=d\psi \w \Pi+d\bar{\psi} \w \bar{\Pi}-L_\D^\pr$ is given by
\bea
H_\D = \Pi \w \f{1}{4}\ga_{ab}\om^{ab}\psi +m\bar{\psi} \psi\Om +\half C_a \bar{\psi}\ga^a\psi \Om. \no
\eea
Although the traditional Hamiltonian density includes $\partial_i \psi$ $(i=1,\cdots,D-1)$, the Hamilton form does not include 
the exterior derivative of the Dirac field.
Rewriting the second and third terms using $\Pi$, we obtain
\bea
H_\D(\psi,\Pi) =  \Pi \w \f{1}{4}\ga_{ab}\om^{ab}\psi +m\Pi \w \p \psi +\half  C_a \Pi \w \theta^a\psi  ,
\eea
with $\p\defe (1/D)\ga_a \theta^a$. 
We have $\ga_a e^a\w \p=\Om$ since \re{A3}.
The Hamilton form is regarded as function only $\psi$ and $\Pi$. 
In the traditional analytic mechanics, the corresponding treatment is equivalent to the formulation using the Dirac bracket. 
However, in the covariant analytic mechanics, the generalization of the Dirac bracket is not known.
In the covariant analytic mechanics, the similar problem does happen for the formulations of the abelian and non-abelian gauge fields and 
of the gravitational field in the second order formalism.

The derivatives of the Hamilton form are given by
\bea
\f{\partial H_\D}{\partial \Pi} \aeq \f{1}{4}\ga_{ab}\om^{ab}\psi +m \p \psi+\half  C_a  \theta^a \psi ,\\
\f{\partial H_\D}{\partial \psi} \aeq \Pi \w \f{1}{4}\ga_{ab}\om^{ab} +m\Pi  \w \p+\half  C_a \Pi \w \theta^a .
\eea
Then, the canonical equation $d\psi = -\partial H_\D/\partial \Pi$ is
\bea
d\psi +\f{1}{4}\ga_{ab}\om^{ab}\psi +m\p\psi+\half  C_a  \theta^a \psi = 0 .
\eea
Applying $\ga_a e^a$ to the above equation from the left and using $\ga_a e^a\w \p=\Om$, we obtain \re{DE}.
The canonical equation $ d\Pi =-\partial H_\D/\partial \psi$ is 
\bea
d\Pi +\Pi \w \f{1}{4}\ga_{ab}\om^{ab} +m\Pi  \w \p+\half  C_a \Pi \w \theta^a = 0 .
\eea
Substituting $\Pi=\bar{\psi} \ga_a e^a$ and using $\Pi \w \p=\bar{\psi} \Om$, \re{trick} and \re{A10}, the above equation becomes \re{DE2}.

\section{Gravity with Dirac field} \la{GD}

We consider the gravitational field coupled with matters including the Dirac field.
We first study the first order formalism and review briefly Nester's approach \cite{N04,N91} in \res{1st}. 
Because the first order formalism is a constrained system, we need to introduce the Lagrange multiplier forms.
Next we study the second order formalism, which is not constrained system.
In \res{L_F}, we investigate the Lagrange formalism, and next, we investigate the Hamilton formalism from \res{H_F} to \res{C_eq}.
The formulations up to \res{H_F} can be easily generalized to $D(\geq 3)$ dimension. 
However, after \res{H_V} we use specialties of $D=4$.

\subsection{First order formalism} \la{1st}

In this subsection, we consider the first order formalism, in which $\theta^a$ and $\om^a_{\ b}$ are independent each other.
The Lagrange form of the gravitational field coupled with the matters is given by
\bea
L^{(1)}(\theta,\om,d\theta,d\om) = L^{(1)}_\G(\theta,\om,d\theta,d\om)+L_\RM{mat}(\theta,\om),
\eea
where $ L^{(1)}_\G$  and $L_\RM{mat}$ are respectively the Lagrange forms of the pure gravity and the matters. The former is given by 
\bea
L^{(1)}_\G = \f{1}{2\ka}\ast R=\f{1}{2\ka}e_{ab}\w( d\om^{ab}+ \om^a_{\ c}\w \om^{cb}), \la{EH}
\eea
with the Einstein constant $\ka=8\pi G/c^3$. 
In \Ref{K}, $L_\RM{mat}$ was $L_\RM{m}(\theta)$ which does not include the connection. 
For instance, $L_\RM{m}(\theta)$ is the  Lagrange form for the scaler field and the abelian and non-abelian gauge fields. 
The variation of $L^{(1)}_\G $ is given by
\bea
\dl L^{(1)}_\G = \f{1}{2\ka}\big[-\dl \theta^c \w e_{abc}\w \Om^{ab}- \dl \om^{ab} \w (\om^c_{\ a} \w e_{cb} +\om^c_{\ b}\w e_{ac})+\dl d\om^{ab} \w e_{ab} \big]. \la{dl_L_G^1}
\eea
Here, we used \re{A5}. We expand the variation of $L_\RM{mat}$ by $\dl \theta^a$ and $\dl \om^{ab}$ as
\bea
\dl L_\RM{mat}(\theta,\om) = -\dl \theta^a \w \ast T_a +\dl \om^{ab} \w \f{\partial L_\RM{mat}}{\partial \om^{ab}} . \la{dl_L_m}
\eea
If we expand $T_a$ as $T_a=T_{ab}\theta^b$, the coefficient $T_{ab}$ is called energy-momentum tensor.
If $L_\RM{mat}=L_\D$, we obtain
\bea
\ast T_c = -\half e_{dc} \w \ga^dD\bar{\psi}\psi+\half e_{dc}\w \bar{\psi} \overleftarrow{D}\ga^d \psi
- e_c m\bar{\psi}\psi, \la{T_c}
\eea
with $D\psi \defe (d\psi+\f{1}{4}\ga_{ab}\om^{ab}\psi)$ and $\bar{\psi} \overleftarrow{D} \defe (d\bar{\psi}-\f{1}{4}\bar{\psi}\ga_{ab}\om^{ab})$.
Applying $\ast$ to the above equation, we obtain
\bea
 T_a \aeq -\half[-\theta_b \ga^b\bar{\psi}D_a\psi+\theta_a \ga^b\bar{\psi}D_b\psi]
+\half [-\theta_b \bar{\psi}\overleftarrow{D}_a\ga^b\psi+\theta_a \bar{\psi}\overleftarrow{D}_b\ga^b\psi]
- \theta_a m\bar{\psi}\psi \no\\
\aeq \half\theta_b \ga^b\bar{\psi}D_a\psi
-\half \theta_b \bar{\psi}\overleftarrow{D}_a\ga^b\psi .
\eea
Here, we used expansions $D\psi= D_a\psi \theta^a$, $\bar{\psi} \overleftarrow{D}=\bar{\psi} \overleftarrow{D}_a \theta^a$ and \re{A2}.
We have $D_a\psi=\theta_a^{\ \mu}(\partial_\mu+\f{1}{4}\ga_{bc}\om^{bc}_{\ \ \mu})\psi$ if we expand the connection as $\om^{ab}=\om^{ab}_{\ \ \mu}dx^\mu$. 
In the second line of the above equation, we used the Dirac equations \re{DE} and \re{DE2}.
We have $T_{ab} = \half \ga_b \bar{\psi}D_a\psi-\half  \bar{\psi}\overleftarrow{D}_a\ga_b\psi$, which coincides the energy-momentum tensor 
derived by the traditional way \cite{Dirac}.
\re{dl_L_G^1} and \re{dl_L_m} lead
\bea
\f{\partial L^{(1)}}{\partial \theta^c}\aeq  -\f{1}{2\ka}e_{abc}\w \Om^{ab} - \ast T_c \com 
\f{\partial L^{(1)}}{\partial d\theta^c}= 0 ,\\
\f{\partial L^{(1)}}{\partial \om^{ab}}\aeq  -\f{1}{2\ka} \big[\om^c_{\ a} \w e_{cb} +\om^c_{\ b}\w e_{ac} \big]+\f{\partial L_\RM{mat}}{\partial \om^{ab}} \com
\f{\partial L^{(1)}}{\partial d\om^{ab}}= \f{1}{2\ka}e_{ab}.
\eea
Then, the Euler-Lagrange equation $\partial L^{(1)}/\partial \om^{ab}+d(\partial L^{(1)}/\partial d\om^{ab})=0$ is
\bea
d e_{ab}-\om^c_{\ a} \w e_{cb} -\om^c_{\ b}\w e_{ac}+2\ka  \f{\partial L_\RM{mat}}{\partial \om^{ab}}  = 0 .\la{de_ab}
\eea
And the Euler-Lagrange equation $\partial L^{(1)}/\partial \theta^c+d(\partial L^{(1)}/\partial d\theta^c)=0$ is 
\bea
-\f{1}{2\ka}e_{abc}\w \Om^{ab} = \ast T_c , \la{E_pre}
\eea
which leads the Einstein equation
\bea
R^a_{\ b} -\half R\dl^a_b = \ka T_b^{\ a}. \no
\eea
We will discuss about \re{de_ab} in \res{L_F}.

The conjugate momentum forms of $\theta^a$ and $\om^{ab}$ are respectively $\pi_a^{(1)}=0$ and $p_{ab}=e_{ab}/2\ka$.
The Hamilton form is
\bea
H^{(1)}(\theta, \om,\pi^{(1)},p) = d\theta^a \w \pi_a^{(1)}+d\om^{ab}\w p_{ab}-L^{(1)}=H_\G^{(1)}-L_\RM{mat} ,
\eea
with
\bea
H_\G^{(1)}= \f{N}{2\ka} \com N\defe e^b_{\ a} \w \om^a_{\ c}\w \om^c_{\ b} .\la{def_N}
\eea
We have $\ast R=e_{ab}\w d\om^{ab}-N$. 
Because this treatment is a constrained system, we need to introduce the Lagrange multiplier forms as 
\bea
H_{\G,\tot}^{(1)}(\theta, \om,\pi^{(1)},p;U,V)= \f{N}{2\ka} +U^a \w \pi_a^{(1)}+V^{ab}\w \big(p_{ab}-\f{e_{ab}}{2\ka}\big) , \la{H_G^1}
\eea
where, $U^a$ and $V^{ab}$ are the Lagrange multiplier forms. 
Using this Hamilton form, we can derive \re{de_ab} and \re{E_pre}. 
In \Ref{IV2}, corresponding treatment of the De Donder-Weyl theory was studied.

The start point of Nester is different from us \cite{N04,N91}. For wide class of the gravitation theories, Nester started from
\bea
L^{(1)}_\G = (d\theta^a+\om^a_{\ b}\w \theta^b)\w \pi_a^{(1)}+(d\om^{ab}+\om^a_{\ c}\w \om^{cb})\w p_{ab}-\Lm(\theta,\om,\pi^{(1)},p) .
\eea
Here, $\Lm$ corresponds to the Hamilton form which is given by a hand depending on the theory. 
So, Nester's approach was not complete Hamilton formalism.
In present theory, $\Lm$ is given by \cite{N04}
\bea
\Lm=U^a \w \pi_a^{(1)}+V^{ab}\w \big(p_{ab}-\f{e_{ab}}{2\ka}\big) ,
\eea
which corresponds to the Lagrange multiplier terms of \re{H_G^1}.

\subsection{Lagrange formalism} \la{L_F}

In the second order formalism, the Lagrange form is different from $L^{(1)}$:
\bea
L(\theta,d\theta) = L_\G(\theta,d\theta)+L_\RM{mat}(\theta,d\theta).
\eea
Here, $L_\G$ is the Lagrange form for the pure gravity given by
\bea
L_\G(\theta,d\theta) = \f{1}{2\ka}N^\pr \com N^\pr\defe \ast R-d(e_{ab}\w \om^{ab}),
\eea
and $L_\RM{mat}(\theta,d\theta)=L_\RM{mat}(\theta,\om(\theta,d\theta))$. 
$\om_{ab}=\om_{ab}(\theta,d\theta)$ is the connection as a function of $\theta^a$ and $d\theta^a$.
The variation  is given by
\bea
\dl L(\theta,d\theta) \aeq -\dl \theta^c \w \Big( \f{1}{2\ka}[e_{abc} \w \Om^{ab}+d(e_{abc}\w \om^{ab})]+\ast T_c \Big)+ \dl d \theta^c \w \f{1}{2\ka}e_{abc}\w \om^{ab} \no\\
&&+ \dl \om^{ab}(\theta,d\theta) \w \Big(\f{1}{2\ka}[de_{ab}-\om^c_{\ a} \w e_{cb} -\om^c_{\ b}\w e_{ac}]+\f{\partial L_\RM{mat}}{\partial \om^{ab}} \Big) .\la{dl_L}
\eea
We suppose that the last term vanishes:
\bea
\f{1}{2\ka}[de_{ab}-\om^c_{\ a} \w e_{cb} -\om^c_{\ b}\w e_{ac}]+\f{\partial L_\RM{mat}}{\partial \om^{ab}} = 0 .\la{de_ab2}
\eea
It is remarkable that this condition is the same with the Euler-Lagrange equation of the connection \re{de_ab} of the first order formalism.
In \Ref{K}, because the Levi-Civit\`{a} connection ($\om^a_{\ b}=w^a_{\ b}$) was supposed and $\partial L_\RM{mat}/\partial \om^{ab}=0$ held since 
$L_\RM{mat}$ was assumed to be independent of the connection, the above requirement \re{de_ab2} was the identity \re{A9}.
\re{de_ab2} is important when $L_\RM{mat}$ includes the Dirac field. 
Under this supposition, \re{dl_L} leads
\bea
\f{\partial L}{\partial \theta^c}=  -\f{1}{2\ka}[e_{abc} \w \Om^{ab}+d(e_{abc}\w \om^{ab})]-\ast T_c \com 
\f{\partial L}{\partial d\theta^c}= \f{1}{2\ka}e_{abc}\w \om^{ab}.
\eea
The Euler-Lagrange equation $\partial L/\partial \theta^c+d(\partial L/\partial d\theta^c)=0$ becomes the Einstein equation \re{E_pre}.

The Lagrange form $L_\G$ is given by subtracting the total differential term $L_\G^\pr\defe \f{1}{2\ka}d(e_{ab}\w \om^{ab})$
from the Einstein-Hilbert form \re{EH}.
The Lagrangian density which correspond to $L_\G^\pr$ is $\f{1}{2\ka}\theta^{-1}\partial_\mu(\theta d^\mu)$ with $d^\mu=2 \theta_{a}^{\ \mu} \theta_b^{\ \lm}\om^{ab}_{\ \ \lm}$.
$\om^{ab}_{\ \ \lm}$ and $d^\mu$ include $d \theta^a$ as \re{om_abc}. 
In the action, the total differential term becomes the surface integration of $\theta d^\mu$.
In the traditional derivation of the Einstein equation, the variation $\dl d^\mu$ is assumed to vanish on the surface even $d^\mu$ contains derivatives of the 
basic variables $\theta^a_{\ \mu}$. 
If we start from $L_\G$, this extra assumption is not needed.
Moreover, we emphasize that it is probably impossible to derive the correct equations from the Einstein-Hilbert form 
including the total differential term $L_\G^\pr$ by the covariant analytic mechanics. 
Probably, it is impossible to expand the variation of $L_\G^\pr$ as
\bea
\dl L_\G^\pr = \dl \theta^a \w X_a+\dl d\theta^a \w Y_a .\no
\eea
If it is possible, $X_a$ and $Y_a$ modify the Euler-Lagrange equation and the conjugate momentum form.
This is very interesting because it suggests that gravitation theories including higher order curvatures ($R^2$, $R^{ab}R_{ab}$, {\it etc}.) can not be treated by the second order formalism 
in the covariant analytic mechanics.

We represent the torsion $C_{abc}$ by the Dirac field.
Using \re{de_ab2} and \re{A9}, we obtain
\bea 
 \f{1}{2\ka }[\tl \om^c_{\ a} \w e_{cb} +\tl \om^c_{\ b}\w e_{ac}]=\f{\partial L_\RM{mat}}{\partial \om^{ab}} \e S_{c,ab}e^c \la{CS}.
\eea
$S_{c,ab}=0$ if $L_\RM{mat}=L_\RM{m}$ and 
\bea
S_{c,ab}=\f{1}{4}\bar{\psi} \half( \ga_c\ga_{ab}+\ga_{ab}\ga_c)\psi=\f{1}{4}\bar{\psi} \ga_{abc}\psi ,\no
\eea
if $L_\RM{mat}=L_\D$ \cite{Dirac}. Here, $\ga_{abc}=\ga_{[a}\ga_b\ga_{c]}$. If $L_\RM{mat}=L_\D^\pr$, $S_{c,ab}=\f{1}{4}\bar{\psi}\ga_c \ga_{ab}\psi$ holds.
The first term in [ ] of \re{CS} becomes $\tl \om^c_{\ a} \w e_{cb} =-C_a  e_b-\tl \om^c_{\ ab}e_c$ using \re{A2} and $\tl \om^c_{\ ac}=-C_a$. 
Similarly, LHS of \re{CS} becomes $-\f{1}{2\ka }[C_a \eta_{cb}-C_{cab}-C_b\eta_{ca}]e^c$ using $2\tl \om_{c[ab]}=-C_{cab}$.
Substituting this to \re{CS}, we obtain
\bea
\f{1}{2\ka }[-C_a \eta_{cb}+C_{cab}+C_b\eta_{ca}] = S_{c,ab}. \la{S_cab}
\eea
Therefore, $C_{cab}$ is represented by the Dirac field. 
Contracting $c$ and $b$ in the above equation, we get
\bea
\f{1}{2\ka }C_a = -\f{S_a}{D-2} ,\la{trC_trS}
\eea
where $D$ is the dimension and $S_a\defe S^b_{\ ab}$. 
Substituting this to \re{S_cab}, we obtain
\bea
\f{1}{2\ka }C_{cab} 
= S_{c,ab}-\f{1}{D-2} [S_a\eta_{cb}-S_b\eta_{ca}]  . \la{C_S}
\eea
If $L_\RM{mat}=L_\D$, $S_a=0$ holds because of the complete anti-symmetric property of $\ga_{abc}$, and we obtain 
\bea
C_a=0, \la{C_a=0}
\eea
from \re{trC_trS}. 
Then, terms including $C_a$ of \res{Dirac field} vanish.
$S_a$ and $C_a $ remain if $L_\RM{mat}=L_\D^\pr$. 
If the Dirac field does not exist ($L_\RM{mat}=L_\RM{m}$), the torsion vanishes.

$N^\pr$ can be rewritten as 
\bea
N^\pr \aeq e_{ab}\w \om^a_{\ c}\w \om^{cb} -de_{ab} \w \om^{ab} \la{def_N2} \no\\
\aeq N+\Theta^a \w  e_{abc}\w \om^{bc}. \la{N^pr_N}
\eea
Substituting \re{de_ab2} to the first line, we obtain $N^\pr= N-2\ka \om^{ab} \w \f{\partial L_\RM{mat}}{\partial \om^{ab}}$. 
Comparing this to the second line, we obtain
\bea
\om^{ab} \w \f{\partial L_\RM{mat}}{\partial \om^{ab}} = -\f{1}{2\ka}\Theta^a \w  e_{abc}\w \om^{bc}. \la{hikaku}
\eea
Kaminaga \cite{K} used $L_\G=\f{1}{2\ka}N$. Because of absence of the Dirac field, this coincides our Lagrange form. 
If the Dirac field exists, $\f{1}{2\ka}N$ is not proper. 
By the way, $N$ can be rewritten as 
\bea
N = d\theta^a \w \half e_{abc}  \w \om^{bc}-\Theta^a \w \half e_{abc}  \w \om^{bc}. \la{N_pi}
\eea
Here, we used $d\theta^a-\Theta^a=-\om^a_{\ b}\w \theta^b $ and $\theta^b \w e_{abc} = -2e_{ca}$ derived from \re{A1} and the definition of $N$, \re{def_N}.

\subsection{Hamilton formalism} \la{H_F}

The conjugate momentum form of $\theta^a$ is given by
\bea
\pi_a=\f{1}{2\ka}e_{abc}\w \om^{bc} \la{def_pi} ,
\eea
and the Hamilton form is given by
\bea
H(\theta,\pi) = d\theta^a \w \pi_a-L=H_\G(\theta,\pi)-L_\RM{mat}(\theta,\pi) ,
\eea
with
\bea
H_\G(\theta,\pi) = \f{N}{2\ka} .
\eea
Here, we used \re{N^pr_N} and \re{N_pi}. Although $L_\G \ne L_\G^{(1)}$, $H_\G=H_\G^{(1)}$ holds. 
In \Ref{K}, $H_\G=L_\G$ was satisfied since $\Theta^a=0$. 
Although the Lagrange form of the pure gravity is different from \Ref{K}, the Hamilton form is the same 
except for that $\om^a_{\ b}$ was the Levi-Civit\`{a} connection in \Ref{K}. 
In \res{H_V}, we represent $N$ by $\theta^a$ and $\pi_a$ and takes derivatives by these.
Since the torsion $C_{abc}$ is represented by the Dirac field, it is independent of $\theta^a$ and $\pi_a$.
Then, $\Theta^a=\half C^a_{\ bc}\theta^b\w \theta^{c}$ is independent of $\pi_a$, however, is a function of $\theta^a$.

The canonical equation for $\theta^a$ is $d\theta^a =\f{1}{2\ka} \f{\partial N}{\partial \pi_a} -\f{\partial L_\RM{mat}}{\partial \pi_a} $. 
In RHS, the second term can be rewritten as
\bea
-\f{\partial L_\RM{mat}}{\partial \pi_a} = -\f{\partial }{\partial \pi_a}\Big[\om^{ab} \w \f{\partial L_\RM{mat}}{\partial \om^{ab}} \Big] 
=\f{\partial }{\partial \pi_a} [\Theta^a \w \pi_a ]= \Theta^a.
\eea
Here, we used \re{hikaku}. Then, the canonical equation becomes
\bea
d\theta^a = \f{1}{2\ka} \f{\partial N}{\partial \pi_a}+\Theta^a \la{H1a}.
\eea
The canonical equation for $\pi_a$ is
\bea
d\pi_a = \f{\partial H}{\partial \theta^a} =\f{1}{2\ka}\f{\partial N}{\partial \theta^a}-\f{\partial L_\RM{mat}}{\partial \theta^a} \la{GOAL}.
\eea
We will show
\bea
\f{\partial N}{\partial\theta^c} = e_{abc} \w \om^{ad} \w \om_d^{\ b}+(\om^d_{\ a}\w e_{dbc}+ \om^d_{\ b}\w e_{adc}+ \om^d_{\ c}\w e_{abd})\w \om^{ab}, \la{Kami}
\eea
in \res{C_eq} using the methods of \Ref{K}.
The above equation is equivalent to 
\bea
\f{\partial N}{\partial\theta^c} = e_{abc} \w \Om^{ab}+d(e_{abc}\w \om^{ab})+2\ka A_{c,ab}\w \om^{ab}, \no
\eea
because of \re{A8}.
Here, $A_{c,ab} \defe \f{1}{2\ka}[\tl \om^d_{\ a}\w e_{dbc}+ \tl \om^d_{\ b}\w e_{adc}+ \tl \om^d_{\ c}\w e_{abd}]$.
Introducing 
\bea
t_c \defe \f{\partial L_\RM{mat}(\theta,\pi)}{\partial\theta^c}+\ast T_c=\f{\partial L_\RM{mat}(\theta,\pi)}{\partial\theta^c} -\f{\partial L_\RM{mat}(\theta,\om)}{\partial\theta^c} ,
\eea
we obtain 
\bea
\f{\partial H}{\partial\theta^c} = +\f{1}{2\ka}[e_{abc} \w \Om^{ab}+d(e_{abc}\w \om^{ab})]+\ast T_c+A_{c,ab}\w \om^{ab}-t_c . \no
\eea
As we show in the remainder of this subsection,
\bea
A_{c,ab}\w \om^{ab}=t_c ,\la{sabun}
\eea
holds. Then, we obtain
\bea
\f{\partial H}{\partial\theta^c} = +\f{1}{2\ka}[e_{abc} \w \Om^{ab}+d(e_{abc}\w \om^{ab})]+\ast T_c. \la{Goal_pre}
\eea
Substituting this to \re{GOAL}, we get the Einstein equation \re{E_pre}.
The RHS of \re{Kami}, which can be rewritten as 
\bea
\om^d_{\ b}\w e_{adc} \w \om^{ab}+\om^d_{\ c}\w e_{abd} \w \om^{ab}\e A_c+B_c, \la{A+B}
\eea
is the same with Sparling's form except for a coefficient and relates to the gravitational energy-momentum pseudo-tensor \cite{N04,N15,S3}.

We show \re{sabun}. Using \re{A1}, $\tl \om^c_{\ ac}=-C_a$ and $2\tl \om_{c[ab]}=-C_{cab}$, we obtain
\bea
A_{c,ab} = \f{1}{2\ka}[2C_{[a} e_{b]c}-C^d_{\ ab}  e_{dc}+2C^d_{\ [a\vert c} e_{d \vert b]}+C_ce_{ab}], \la{A}
\eea
with $2C^d_{\ [a\vert c} e_{d \vert b]}=C^d_{\ ac} e_{d b}-C^d_{\ bc} e_{d a}$.
By the way, using \re{hikaku}, \re{A6} and \re{A7}, we obtain
\bea
\f{\partial L_\D}{\partial\theta^c} = \half \bar{\psi}\ga_a  e^a_{\ c} \w d\psi-\half e^a_{\ c} \w d\bar{\psi}\ga_a\psi 
+ e_c m\bar{\psi}\psi - C^a_{\ cb}  \theta^b \w \pi_a . \la{it}
\eea
Then, for $L_\RM{mat}=L_\D$, we get
\bea
t_c = -e_{dc}S^d_{\ ab}\w \om^{ab}- C^a_{\ cb}  \theta^b \w \pi_a , \la{t_c}
\eea
using \re{it} and \re{T_c}. 
This relation also holds for $L_\RM{mat}=L_\D^\pr$. We have $t_c=0$ if $L_\RM{mat}=L_\RM{m}$. 
\re{t_c} can be rewritten as $t_c=B_{c,ab}\w \om^{ab}$ with
\bea
B_{c,ab}=-e_{dc}S^d_{\ ab}+ \f{1}{2\ka}[C_c e_{ab}-C^d_{\ c[a\vert } e_{d \vert b]}]. \la{mA}
\eea
Here, we used \re{def_pi} and \re{A1}. We can show $A_{c,ab}=B_{c,ab}$ using \re{A}, \re{mA} and \re{C_S}.
Then, \re{sabun} holds.

\subsection{Variation of the Hamilton form} \la{H_V}

We represent $\om^{ab}$ and $N$ by $\pi_a$. In $D$ dimension space-time, $e_{abc}$ is given by
\bea
e_{abc} = \f{1}{(D-3)!}E_{d_1 \cdots d_{D-3}abc}\theta^{d_1} \w \cdots \w \theta^{d_{D-3}}, \no
\eea
where $E_{d_1 \cdots d_D}$ is the complete anti-symmetric tensor such that $E_{01\cdots D-1}=1$.
In the following, we use specialties of $D=4$. 
Substituting $e_{abc}=E_{dabc}\theta^d$ and $\om^{ab}= \om^{ab}_{\ \ c}\theta^c$ to \re{def_pi}, we obtain
\bea
\pi_c =\half \pi_{c,ab}\theta^a \w \theta^b \com 
\pi_{c,ab} = \f{1}{2\ka}(E_{adec}\om^{de}_{\ \ b}-E_{bdec}\om^{de}_{\ \ a}) \la{pi_cab}.
\eea
Using the technique used to get \re{om_abc} from \re{FS} (this technique can not be used for $D \ne 4$), we obtain
$\f{1}{2\ka}E_{bdea}\om^{de}_{\ \ c}=\half (\pi_{b,ca}+\pi_{c,ba}+\pi_{a,bc})$. 
It leads
\bea
\om^{ab}_{\ \ c} = \f{\ka}{4}E^{abnm}\tau_{mcn} \com
\tau_{abc} \defe \pi_{b,ac}+\pi_{a,bc}+\pi_{c,ab} \la{om_t}.
\eea
We have $\tau_{abc}=-\tau_{cba}$. $\tau_{abc}$ is represented by $\pi_c$ as $\tau_{abc}=-\ast p_{abc}$ with
\bea
p_{abc} \defe \pi_b \w e_{ac}+\pi_a \w e_{bc}+\pi_c \w e_{ab} .
\eea
Here, we used \re{A4}. Substituting this equation to \re{om_t}, we obtain
\bea
\om^{ab}= -\f{\ka}{4}E^{abnm}\ast p_{mcn}\theta^c . \la{om_p} 
\eea
This equation and \re{FS} lead
\bea
d\theta^a -\Theta^a =\f{\ka}{4}E^{abnm}\ast p_{mcn}\theta^c \w \theta_b .\la{kouzou}
\eea
Substituting $\pi_{c,ab}=-\ast p_{[ab]c}$ to the first equation of \re{pi_cab}, we obtain
\bea
\pi_c = -\half \ast p_{cab} \theta^a \w \theta^b . \la{pi_p}
\eea

We calculate the derivatives of $N$ by $\theta^a$ and $\pi_a$.  Substituting \re{om_p} to \re{def_N}, we obtain \cite{K}
\bea
N = n^{a_1a_2a_3a_4a_5a_6}\ast p_{a_1a_2a_3}\ast p_{a_4a_5a_6}\Om ,
\eea
by using \re{A4}. Here, 
\bea
n^{a_1a_2a_3a_4a_5a_6} \aeq \f{\ka^2}{16}[ \eta^{a_2a_6}\eta^{a_3a_5}\eta^{a_1a_4}+\eta^{a_2a_4}\eta^{a_3a_6}\eta^{a_1a_5}
+\eta^{a_5a_6} \eta^{a_3a_4} \eta^{a_1a_2}+\eta^{a_5a_4}\eta^{a_3a_2}\eta^{a_1a_6} \no\\
&&-\eta^{a_2a_6}\eta^{a_3a_4}\eta^{a_1a_5}-\eta^{a_2a_4}\eta^{a_3a_5}\eta^{a_1a_6}
-\eta^{a_5a_6} \eta^{a_3a_2} \eta^{a_1a_4}-\eta^{a_5a_4}\eta^{a_3a_6}\eta^{a_1a_2}], \la{n^123456}
\eea
of which symmetries are
\bea
n^{a_1a_2a_3a_4a_5a_6}=-n^{a_3a_2a_1a_4a_5a_6}=-n^{a_1a_2a_3a_6a_5a_4}=n^{a_4a_5a_6a_1a_2a_3}. \la{sym}
\eea
Using the formula
\bea
\dl \ast p_{abc} \xi = (\dl p_{abc}+\dl \Om \ast p_{abc})\ast \xi, \no
\eea
for the arbitrary 4-form $\xi$ and \re{sym}, \re{A5} and \re{A7}, we obtain \cite{K}
\bea
\f{\partial N}{\partial \pi_a} \aeq -2n^{dbcnml}(2\dl^a_d e_{bc}+ \dl^a_b e_{dc}) \ast p_{nml} , \la{dH_dpi}\\
\f{\partial N}{\partial\theta^a} \aeq n^{dbcnml}[4\pi_{d}\w e_{bca}+2\pi_{b}\w e_{dca}+\ast p_{dbc} e_a] \ast p_{nml}. \la{dH_dtheta}
\eea

\subsection{Canonical equations} \la{C_eq}

We calculate RHS of \re{H1a}. 
Substituting \re{n^123456} to $n^{dbcnml}(2\dl^a_d e_{bc}+ \dl^a_b e_{dc})$ of RHS of \re{dH_dpi}, and using \re{A1}, 
we obtain
\bea
n^{dbcnml}(2\dl^a_d e_{bc}+ \dl^a_b e_{dc}) = \f{\ka^2}{4}\theta^m \w e^{lna} .\no
\eea
Substituting this equation to \re{dH_dpi} and using $e^{lna}=E^{blna}\theta_b$, we obtain
\bea
\f{1}{2\ka}\f{\partial N}{\partial \pi_a} 
= \f{\ka}{4}E^{a b ln}\ast p_{nml} \theta^m \w  \theta_b .
\eea 
Then, the canonical equation for $\theta^a$ \re{H1a} becomes
\bea
d\theta^a = \f{\ka}{4}E^{a b ln}\ast p_{nml} \theta^m \w  \theta_b +\Theta^a ,
\eea
which coincides with \re{kouzou}. 
The above equation and \re{FS} lead \re{om_p}, which is equivalent to the definition of the conjugate momentum form $\pi_c$.

We show \re{Kami}. Substituting \re{pi_p} to RHS of \re{dH_dtheta}, and using the symmetry $p_{abc}=-p_{cba}$ and \re{A14}, 
we can obtain \cite{K}
\bea
\f{\partial N}{\partial\theta^c} = -
 n^{abdnml}
\big[2(\ast p_{dbc}+\ast p_{bdc})e_a+2(\ast p_{adc}-\ast p_{acd})e_b+\ast p_{abd}e_c\big] \ast p_{nml}. \no
\eea
Substituting \re{n^123456} to this equation, we obtain
\bea
\f{\partial N}{\partial\theta^c} \aeq \f{\ka^2}{2}e_n(\ast p^{[ab]n}\ast p_{acb}+\ast p_{ac}^{\ \ n}\ast p^{a}) \no\\
 && +\f{\ka^2}{2}e_n( - \ast p_{abc}\ast p^{n a b}+\ast p^n_{\ ac}\ast p^a_{\ }-\ast p_{c}\ast p^n) \no\\
&&+\f{\ka^2}{4}e_c( -\ast p_{abd}\ast p^{b[ad]}+\ast p_a\ast p^a), \la{tikara}
\eea
with $p_a=p^b_{\ ba}$. In \Ref{K}, the anti-symmetrization symbols were missed. 

Next, we show that RHS of the above equation becomes RHS of \re{Kami}, namely \re{A+B}.
Substituting \re{om_p} to $A_c=\om^{ab}\w \om^d_{\ b} \w e_{adc} $, we obtain $ A_c^{(1)}+A_c^{(2)}$ with
\bea
A_c^{(1)} \aeq \f{\ka^2}{16}E^{abnm}E^{d\ kl}_{\ b}[(\dl^s_d\dl^t_c-\dl^s_c\dl^t_d)e_a+(\dl^s_c\dl^t_a-\dl^s_a\dl^t_c )e_d]\ast p_{msn}  \ast p_{ltk}  \no\\
\aeq \f{\ka^2}{8}e_aE^{abnm}E^{d\ kl}_{\ b}(\dl^s_d\dl^t_c-\dl^s_c\dl^t_d) \ast p_{msn}  \ast p_{ltk} ,\\
A_c^{(2)} \aeq \f{\ka^2}{16}e_cE^{abnm}E^{d\ kl}_{\ b}( \dl^s_a\dl^t_d- \dl^s_d\dl^t_a) \ast p_{msn}  \ast p_{ltk} .
\eea
Here, we used \re{A14}. We can show that
\bea
A_c^{(1)} \aeq \f{\ka^2}{2}e_n(-\ast p^{[ab]n}\ast p_{acb}-\ast p_{ac}^{\ \ n}\ast p^{a}) ,\la{A_c1}\\
A_c^{(2)} \aeq \f{\ka^2}{4}e_c (\ast p_{abd}  \ast p^{b[ad]}-\ast p_{a}  \ast p^{a}). \la{A_c2}
\eea
Similarly, we can write $B_c= \om^{ab}\w \om^d_{\ c}\w e_{abd}$ as $ B_c^{(1)}+B_c^{(2)}$ with
\bea
B_c^{(1)} \aeq \f{\ka^2}{16}E^{abnm}E_{\ c}^{d\ kl}[(\dl^s_b\dl^t_d-\dl^s_d\dl^t_b)e_a+(\dl^s_d\dl^t_a-\dl^s_a\dl^t_d )e_b] \ast p_{msn}  \ast p_{ltk} \no\\
\aeq \f{\ka^2}{8}e_aE^{abnm}E_{\ c}^{d\ kl}(\dl^s_b\dl^t_d-\dl^s_d\dl^t_b) \ast p_{msn}  \ast p_{ltk} ,\\
B_c^{(2)} \aeq \f{\ka^2}{16}E^{abnm}E_{\ c}^{d\ kl} \ast p_{msn}  \ast p_{ltk} ( \dl^s_a\dl^t_b- \dl^s_b\dl^t_a)e_d \no\\
\aeq \f{\ka^2}{8}e_d E^{abnm}E_{\ c}^{d\ kl} \ast p_{man}  \ast p_{lbk}.
\eea
And these equations lead, 
\bea
B_c^{(1)} \aeq  \f{\ka^2}{2}e_n(\ast p^{[ab]n}\ast p_{acb}+\ast p_{ac}^{\ \ n}\ast p^{a}) \no\\
&&+\f{\ka^2}{2}e_n(\ast p_{[ab]c}\ast p^{anb}
-\ast p_{abc}\ast p^{nab}+\ast p^{\ n}_{a \ c}\ast p^a-\ast p_c\ast p^n) \no\\
&&+\f{\ka^2}{2}e_c(-\ast p_{abd}\ast p^{b[ad]}+ \ast p_a\ast p^a),\la{B_c1}\\
B_c^{(2)} \aeq \f{\ka^2}{2}e_n(\ast p^{[ab]n}\ast p_{acb}+\ast p_{ac}^{\ \ n}\ast p^{a}) \no\\
&&+\f{\ka^2}{2}e_n(-\ast p_{[ab]c}\ast p^{anb}-\ast p^{\ n}_{a \ c}\ast p^a +\ast p^{n}_{\ a c}\ast p^{a}). \la{B_c2}
\eea
Using \re{A_c1}, \re{A_c2}, \re{B_c1} and \re{B_c2} (the anti-symmetrization symbols were missed in \Ref{K}), 
RHS of \re{tikara} becomes RHS of \re{Kami}, namely \re{A+B}. 
Therefore, we obtain \re{Goal_pre}, and
the canonical equation for $\theta^a$ \re{GOAL} becomes the Einstein equation \re{E_pre}.

\section{Summary}

We applied the covariant analytic mechanics with the differential forms to the Dirac field and the gravity with the Dirac field.
In \res{CM}, we reviewed the covariant analytic mechanics which treats space and time on an equal footing regarding the differential forms as the basis variables 
and has significant advantages that the canonical equations are gauge covariant as well as manifestly diffeomorphism covariant. 
Combining our study and the previous works \cite{Na,K,N04} (the scalar field, the Proca field, the electromagnetic field, the non-abelian gauge field 
and the gravity without the Dirac field), the applicability of the covariant analytic mechanics was checked for all fundamental fields. 

In \res{Dirac field}, we studied the Dirac field. 
The frame (vielbein) is necessary to write down the Lagrange form even in the flat space-time. 
This fact represents a nature of the Dirac field. 
We regarded the basis variable of the Hamilton form of the Dirac field as only $\psi$ and its conjugate momentum form $\Pi$. 
In the traditional analytic mechanics, the corresponding treatment is equivalent to the formulation using the Dirac bracket. 
In the covariant analytic mechanics, the similar problem does happen for the formulation of other fundamental fields. 
Although we introduced the Poisson bracket of the covariant analytic mechanics for the first time, 
the possibilities of applications to the Dirac bracket, the canonical transform theory and the quantization are unclear. 

In \res{GD}, we studied gravitational field coupled with matters including the Dirac field and claimed that 
Nester's approach \cite{N04,N91} was not complete Hamilton formalism. 
Although the second order formalism is not constrained system, the first order formalism is a constrained system, which needs the Lagrange multiplier forms.
In the second order formalism, the Lagrange form of the pure gravity is given by subtracting the total differential term 
$L_\G^\pr=\f{1}{2\ka}d(e_{ab}\w \om^{ab})$ from the Einstein-Hilbert form. 
If we do not drop the $L_\G^\pr$, it is probably impossible to derive the correct equations.
This is very interesting because it suggests that gravitation theories including higher order curvatures can not be treated by the second order formalism 
in the covariant analytic mechanics. 
The torsion was determined by the condition that the last term of RHS of \re{dl_L} vanishes. 
Although the Lagrange form of the pure gravity was different from \Ref{K}, the Hamilton form was the same 
except for that the connection was the Levi-Civit\`{a} connection in \Ref{K}. 
We took the derivatives of the Hamilton form of the pure gravity using the specialties of 4 dimension system and corrected the 
errors of \Ref{K}. 
In other part, which can be easily generalized to arbitrary dimension, we treated the contributions due to the Dirac field.

\acknowledgments

We acknowledge helpful discussions with Y. Tokura and Y. Kaminaga.

\appendix

\section{Formulas} \la{formula}

Several useful Formulas are listed. For $\theta^a \w e_{a_1 \cdots a_r}(r=1,2,3)$,
\bea
(-1)^D\theta^a \w e_{bcd} \aeq -\dl^a_b e_{cd}+\dl^a_c e_{bd}-\dl^a_d e_{bc} \la{A1},\\
(-1)^D\theta^a \w e_{bc} \aeq \dl^a_b e_{c}-\dl^a_c e_{b}  \la{A2},\\
(-1)^D\theta^a \w e_b \aeq -\dl^a_b \Om , \la{A3}
\eea
hold \cite{S3}. Using \re{A2} and \re{A3}, we obtain
\bea
\theta^a \w \theta^b \w e_{cd} = (\dl^a_c\dl^b_d-\dl^a_d\dl^b_c)\Om. \la{A4}
\eea
 Using \re{A1} and \re{A2}, we have
\bea
\theta^a \w \theta^b\w e_{cde} =(\dl^a_d\dl^b_e-\dl^a_e\dl^b_d)e_c+(\dl^a_e\dl^b_c-\dl^a_c\dl^b_e )e_d+( \dl^a_c\dl^b_d- \dl^a_d\dl^b_c)e_e. \la{A14}
\eea
For $\dl e_{a_1 \cdots a_r}(r=0,1,2)$, 
\bea
\dl e_{ab} \aeq -(-1)^D\dl \theta^c \w e_{abc} ,\la{A5}\\
\dl e_a \aeq -(-1)^D\dl \theta^b \w e_{ab} \la{A6} ,\\
\dl \Om \aeq -(-1)^D\dl \theta^a \w e_a, \la{A7}
\eea
hold. 
For $de_{a_1 \cdots a_r}(r=1,2,3)$, we have \cite{S3}
\bea
d e_{abc}\aeq  w^d_{\ a}\w e_{dbc}+ w^d_{\ b}\w e_{adc}+ w^d_{\ c}\w e_{abd} \la{A8} ,\\
d e_{ab}\aeq w^c_{\ a} \w e_{cb} +w^c_{\ b}\w e_{ac} \la{A9},\\
d e_a\aeq w^b_{\ a} \w e_b =-(\om_a+C_a)\Om. \la{A10}
\eea

\end{document}